\documentstyle[11pt,aaspp4]{article}

\lefthead{White et al.}
\righthead{WBL Poor Cluster Catalog}

\begin{document}

\title{A Catalog of Nearby Poor Clusters of Galaxies}

\author{Richard A. White}
\affil{NASA Goddard Space Flight Center, Code 631, Greenbelt, MD 20771}

\author{Mark Bliton\altaffilmark{1}}
\affil{Department of Astronomy, New Mexico State University, Las Cruces, NM 88003}

\author{Suketu P. Bhavsar}
\affil{Department of Physics \& Astronomy, University of Kentucky, Lexington, KY 40506}

\author{Patricia Bornmann}
\affil{NOAA Space Environment Center, Boulder, CO 80303}

\author{Jack O. Burns\altaffilmark{2}}
\affil{Department of Physics \& Astronomy, University of Missouri, Columbia, MO 65211}

\author{Michael J. Ledlow}
\affil{Department of Physics \& Astronomy, University of New Mexico, Albuquerque, NM 87131}

\and

\author{Christen Loken}
\affil{Department of Physics \& Astronomy, University of Missouri, Columbia, MO 65211}

\altaffiltext{1}{mailing address:  Department of Physics \& Astronomy, University of Missouri, Columbia, MO 65211}
\altaffiltext{2}{Office of Research, University of Missouri, Columbia, MO 65211}

\begin{abstract}
A catalog of 732 optically selected, nearby poor clusters of galaxies
covering the entire sky north of $\rm -3^{\circ}$ declination is
presented.  The poor clusters, called WBL clusters, were identified as
concentrations of 3 or more galaxies with photographic magnitudes
brighter than 15.7, possessing a galaxy surface overdensity of
$10^{4/3}$. These criteria are consistent with those used in the
identification of the original Yerkes poor clusters, and this new
catalog substantially increases the sample size of such objects.
These poor clusters cover the entire range of galaxy associations up
to and including Abell clusters, systematically including poor and rich
galaxy systems spanning over three orders of magnitude in the cluster
mass function.  As a result, this new catalog contains a greater
diversity of richness and structures than other group catalogs, such
as the Hickson or Yerkes catalogs.  The information on individual
galaxies includes redshifts and cross-references to other galaxy
catalogs.  The entries for the clusters include redshift (where
available) and cross-references to other group and cluster catalogs.
\end{abstract}

\keywords{galaxies: clusters: general --- catalogs}

\section{Introduction}
Traditionally, the study of galaxy associations has focused on rich
clusters of galaxies such as the Abell clusters (\cite{abe58}; Abell,
Corwin, \& Olowin 1989).  However, these clusters represent the
extreme in galaxy associations, and although massive, are relatively
rare.  Conversely, poor clusters are less massive but more numerous,
and constitute a significant fraction of the mass of the universe.  As
the building blocks of clusters and superclusters, poorer galaxy
associations must be studied in order to understand the formation and
evolution of large-scale structure in the Universe. Poor clusters are
fundamental entities, but until recently, they have not received as
much attention as they merit.  This stems in part from a paucity of
catalogs that reach, in a consistent way, out to Abell cluster
distances.

There are several poor cluster catalogs in the literature which have
yielded important scientific results.  The CfA group catalog
(\cite{gel83}), based on both redshift and spatial information
(thereby removing the possibility of projection effects) showed groups
to be important in tracing large-scale structure.  The Hickson Compact
Group catalog (\cite{hic82}; HCG), with its high galaxy density
criterion, has proved an important resource for studying interacting
galaxies.  The Albert, White, \& Morgan (1977; AWM), Morgan, Kayser,
\& White (1975; MKW) and White (1978; WP) clusters, also known as
Yerkes clusters, specifically targeted poor clusters containing cD
galaxies.  These clusters have been studied extensively in the
optical, radio, and x-ray regions of the spectrum.  They possess
optical structures that are a continuation of the Abell clusters to
lower richness levels (\cite{bah80}), and their x-ray and radio
properties show surprising similarities to the Abell clusters as well
(e.g. x-ray cooling flows and tailed radio sources---\cite{doe95};
\cite{bur87}).

The original identifications of the Yerkes clusters were made by eye
in a laborious procedure of scanning the POSS glass plates in
conjunction with the Zwicky catalog (\cite{zwi61}; CGCG).  The new
study presented in this paper was undertaken to create a much larger
sample of clusters electronically (with galaxy overdensities similar
to the Yerkes clusters), utilizing an algorithm that reproduced the
human eye search.

The terms ``poor galaxy cluster'' and ``galaxy group'' have not been
consistently defined or applied in the literature.  Since this catalog
includes galaxy associations of all richness levels (see
e.g. \cite{bha81}), we use the term ``poor cluster'' to describe these
entries.  Imposing an arbitrary change in nomenclature, calling rich
associations clusters and poor associations groups would obscure the
continuous spectrum of properties possessed by these objects.

\section{Techniques}

The creation of the catalog employed the Turner \& Gott (1976; TG)
algorithm which is capable of mimicking, in a very mathematical,
reproducible, and consistent way, the work of the human eye in picking
out density enhancements from the POSS.  We chose to apply the TG
algorithm to the CGCG catalog, since this procedure accomplished our
goal of identifying the majority of the Yerkes clusters.  The CGCG is
a compilation of galaxies and clusters, covering the sky north of -3
degrees declination.  Zwicky identified galaxies to a limiting
photographic magnitude, $m_{pg}=15.7$ on the POSS photographic
plates. Zwicky clusters were identified as galaxy associations
containing at least 50 galaxies in the magnitude range $m_{max}$ to
$m_{max}+3$, where $m_{max}$ is the magnitude of the brightest cluster
member.  To create this poor cluster catalog, we used CGCG galaxies as
faint as $m_{pg}$=15.7, which is 0.7 magnitudes deeper than previous
poor cluster searches (e.g. Bhavsar 1980).

The TG algorithm begins with an assigned target factor for galaxy
surface density enhancement ($\sigma_g$). Around each galaxy in the
catalog it determines the smallest circular aperture on the sky, centered
on the galaxy, that yields the desired $\sigma_g$.  The number of
additional galaxies, if any, within each aperture is also noted, which
we define to be the number of ``nearest neighbors.''  All overlapping
apertures are then merged together into a cluster.  Galaxies that
have no nearest neighbors and are not included within the apertures of
other galaxies are considered isolated.  The process is then repeated,
at a larger value of $\sigma_g$, excluding all isolated galaxies.
This results in a hierarchy of clusters which shows a given cluster's
fragmentation or continued integrity with increasing $\sigma_g$.  A
more detailed discussion of the algorithm and its operation may be
found in earlier papers (TG; \cite{bha83}; \cite{bha80}).

Levels of $\sigma_g$ were chosen to be multiplicative factors above
the average surface density of CGCG galaxies determined in the region
of the galactic caps ($|b|>40^{\circ}, \ \delta > 0^{\circ}$; 6866
galaxies/steradian).  Values of $\sigma_g$ were increased in intervals
of $10^{1/3}$ (beginning with $10^{2/3}$) which corresponds nominally
to a volume density enhancement, $\rho_g$ = $10^{1/2}$ (beginning with
10).

In order to reduce the role of chance line-of-sight galaxy projections
creating false clusters in the catalog, we sought an enhancement level
low enough to select all the MKW, AWM, and WP clusters, but no lower.
The two most optimum levels were found to be $\sigma_g = 10^{4/3}
\approx 21$ ($\sigma_{21}$) and $\sigma_g = 10^{5/3} \approx 46$
($\sigma_{46}$), corresponding to $\rho_g$ = 100 and 316,
respectively.  The $\sigma_{21}$ results were used to identify the
clusters, and the $\sigma_{46}$ information was used both to detect
regions of hierarchical subclustering and to identify a subsample of
poor clusters at a higher density.  Subsequent analysis has shown that
projection effects play a negligible role in these poor clusters
(\cite{bur96}; \cite{bha80}).  The number of poor clusters detected at
$\sigma_{21}$ and $\sigma_{46}$, along with the total number of
galaxies which are members of these poor clusters, are listed in Table
1.  The typical aperture radius of a 3 member cluster at $\sigma_{21}$
is 0.146 degrees; at $\sigma_{46}$ it is 0.099 degrees.  To excellent
approximation, for richer clusters, the aperture size scales with the
square root of the number of members.

\begin{table}
\dummytable\label{tbl-1}
\end{table}

\section{The Poor Cluster Catalog}

The following selection criteria were used to create the full poor cluster 
catalog:
$$\sigma_{g} \geq 10^{4/3} = \sigma_{21}$$
$$m_{pg} \leq 15.7$$
$$N_g \geq 3,$$
where $\sigma_{g}$ is the surface density of galaxies above the background,
as defined in section 2, $ m_{pg}$ is the photographic magnitude of
each galaxy as recorded in the CGCG, and $N_g$ is the number of CGCG
galaxies in each cluster.  This catalog is presented in Table 2.

The electronic version of the CGCG we used has 15409
galaxies within the galactic caps.  In this region, the WBL
clusters contain 2245 galaxies at $\sigma_{21}$ and 1159 galaxies at
$\sigma_{46}$.  The total area of this region is 2.24431 steradians.
The area occupied by WBL clusters in this region is 0.01416
steradians at $\sigma_{21}$ and 0.00346 steradians at
$\sigma_{46}$.  Thus the WBL clusters contain about 14.6\% of the
galaxies in 0.63\% of the area at $\sigma_{21}$ and 7.5\% of the
galaxies in 0.15\% of the area at $\sigma_{46}$.  These results make
clear that this catalog contains poor clusters of high galaxy density.

The summary data for each poor cluster are presented in Table 2, and
the data on individual galaxy members in Table 3.  Both Tables 2 and 3
will be available in electronic form from the Astronomical Data Center
({\it http://adc.gsfc.nasa.gov/}).  Table 4 contains a cross reference
to previously used names for these poor clusters.  The new clusters
are designated ``WBL'' to represent the last names of the authors,
which is consistent with IAU guidelines.

\subsection{The Poor Cluster Catalog:  Table 2}

This table presents basic data on all 732 poor clusters.  The
following information is provided:
  
Column 1: Identification.  The WBL designation of each poor cluster.
Cross-references to previously used names are given in Table 4.
Footnotes in this column provide additional information for the poor
clusters where necessary.

Column 2: Coordinates.  The right ascension and declination, equinox
B1950, for the centroid of the poor cluster are listed.  The centroid
is determined from a luminosity weighted mean of all member galaxies
listed in the CGCG.  Luminosities were computed as $L =
10^{-0.4m_{pg}}$, and used as the weight in computing the mean RA and
DEC for each cluster, i.e.: $\langle RA \rangle = {\sum_i{(RA)_i \
L_i} \over {\sum_i{L_i}}}.$

Column 3: Richness.  The number of CGCG galaxies in each poor
cluster.

Column 4: Clustering at $\sigma_{46}$.  An indication of the fate of
each individual galaxy at the higher density enhancement
($\sigma_{46}$).  A single 0 in this column indicates that a group at
$\sigma_{21}$ fractured into single galaxies at $\sigma_{46}$
(i.e. the galaxy apertures did not overlap).  Combinations of other
numbers indicate how many galaxies were in each subcluster at
$\sigma_{46}$, with 0 indicating one or more isolated galaxies.  For
example, a cluster with 11 members and an 8+0 in this column becomes a
group of 8 members with 3 isolated galaxies at $\sigma_{46}$.  A
cluster of 11 members and an entry of 4+3+2+0 breaks up into 3
subclusters of 4, 3, and 2 members, with 2 isolated galaxies at
$\sigma_{46}$.

Columns 5-7: Poor cluster redshift.  Column 5 lists a redshift for the
cluster, when available, computed as an average of redshifts from the
literature obtained through the NASA Extragalactic Database
(NED)\footnote{The NASA/IPAC Extragalactic Database (NED) is operated
by the Jet Propulsion Laboratory, California Institute of Technology,
under contract with the National Aeronautics and Space
Administration.}.  Column 6 ($N_z$) indicates the number of galaxy
redshifts available from NED.  Column 7 (Note) gives notes on the
determination of the cluster redshift.  If there is no note in column
7, the number of galaxy redshifts listed in column 6 were used to
compute the cluster redshift in column 5.  If the note is $i$, there
were only two redshifts available and their values differed by more
than 1500 $km \ s^{-1}$.  In this case, the two galaxies may be close
only in projection. The redshift given in this case is the average of
the two galaxy redshifts.  A note of $ii$ indicates that, of the 3 or
more redshifts available, one was more than 1500 $km \ s^{-1}$ from
the mean.  After removing this discrepant redshift, a new mean was
calculated, with all remaining redshifts within 1500 $km \ s^{-1}$
from this new mean.  A note of $iii$ indicates that, of the 3 or more
redshifts available, at least two discrepant redshifts were present.
The redshift listed in column 7 is then the mean redshift calculated
after removing only the most discrepant velocity.  A value of 1500 $km
\ s^{-1}$ was chosen for the cutoff velocity because a ``typical''
poor cluster has a velocity dispersion, $\sigma_v$, $\approx$ 500 $km
\ s^{-1}$ (\cite{led96}; \cite{bee95}).  This criteria is thus
analogous to the standard 3$\sigma$ clipping often used in the
literature.

Columns 8-10: Zwicky cluster cross-correlation.  Column 8 lists the
Zwicky cluster containing the given WBL cluster within its contours.
Column 9 ($R_Z$) gives the radius of the Zwicky cluster in arcmin and
column 10 ($f_Z$)lists the distance separating the center of the
Zwicky and WBL clusters, in units of the Zwicky cluster radius.  The
radius is that given in the CGCG, and is meant to indicate the area on
the sky contained within the hand-drawn contours.  Zwicky clusters are
rarely spherical, so all correlations were verified by eye.  Zwicky
clusters are also separated into estimated distance classes, with near
clusters meant to be in the redshift range $0.0 < z \leq 0.05$, and
medium distant clusters in the range $0.05 < z \leq 0.1$ (CGCG).  We
included any near or medium distant Zwicky cluster which contained a
WBL cluster within its contours.  Near clusters were chosen over
medium distant clusters in the very few cases where a poor clusters
fell within the boundaries of two Zwicky clusters of different
distance classes.  The letters MD attached to a Zwicky cluster name 
(column 8) indicate a medium distant cluster.

Column 11: Abell cluster cross-correlation.  Any WBL cluster located
within one corrected Abell radius (\cite{abe89}) 
of a distance class 3 or nearer
Abell cluster was considered associated with the Abell cluster.
Additionally, NED was used to search for WBL
cluster members that were also members of Abell clusters.

Column 12: CfA group cross-correlation.  Associated CfA groups were
identified as groups whose coordinates were within 30 arcmin of the
WBL centroid.

Column 13: HCG cross-correlation.  Associated Hickson compact groups
were identified from information on individual galaxies obtained through NED.

Column 14: Yerkes cluster cross-correlation.  The identification of
WBL clusters associated with the Yerkes clusters was determined
visually as a part of the development of the WBL catalog.

\begin{table}
\dummytable\label{tbl-2}
\end{table}

\subsection{Poor Cluster Galaxies:  Table 3}

This table presents basic data on each Zwicky galaxy included as a
poor cluster member.  Due to the length of this table, it will only be
available electronically.  The following information is provided:

Column1:  Identification.  The WBL cluster to which the galaxy
belongs.  

Columns 2-3: Coordinates.  The right ascension and declination of the
member galaxy from the CGCG in B1950 coordinates.

Column 4: Magnitude.  The apparent photographic magnitude, $m_{pg}$,
of the galaxy from the CGCG.

Columns 5-6: Nearest neighbors.  A measure of the number of nearest
neighbors to the galaxy at $\sigma_{21}$ ($N_{21}$, column 5), and
$\sigma_{46}$ ($N_{46}$, column 6).  This refers to the number of
neighbor galaxies falling within the aperture defining the
$\sigma_{21}$ or $\sigma_{46}$ threshold.  For a galaxy with no
nearest neighbor, the aperture radius is 0.084 degrees at
$\sigma_{21}$ and 0.057 degrees at $\sigma_{46}$.  The radius scales
as $\sqrt{n+1}$, where $n$ is the number of nearest neighbors.  For an
individual galaxy, the higher the number, the more it is centrally
located within the poor cluster.  Multiple poor cluster members with
high numbers of neighbors indicate a compact cluster.  Galaxies with a
nearest neighbor number of zero were merged into a cluster because a
nearby galaxy possessed many neighbors.  The resulting large aperture
($\propto \sqrt{n+1}$) may have overlapped an isolated galaxy in some
cases.

Column 7: Clustering at $\sigma_{46}$.  An indication of the fate of
each individual galaxy at the higher density enhancement.  A blank field
indicates the galaxy became isolated (no neighbors and no overlapping
apertures), and therefore was not considered a member of a cluster at
$\sigma_{46}$.  Galaxies with the same letters are part of the same
poor cluster at $\sigma_{46}$.  WBL designations for these subgroups
should include this letter and indicate that it is a $\sigma_{46}$
cluster.

Column 8: Galaxy redshift.  The redshift of the galaxy reported in
NED.  Several Zwicky galaxies are actually multiple galaxies, and
therefore have multiple identifications in NED.  In these instances,
the average of all redshifts available for the Zwicky galaxy are
presented.

Column 9: Galaxy catalog cross-correlations.  Cross-identifications
for the galaxies from the NGC (\cite{dre88}), UGC (\cite{nil73}), and
IC (\cite{dre88}) catalogs obtained from NED.  For entries that are
actually multiple galaxies, all relevant identifications are
presented.

\begin{table}
\dummytable\label{tbl-3}
\end{table}

\subsection{Previous Nomenclature:  Table 4}

This table cross-references the names of the poor clusters presented
here with previous names used in the literature.  We also present a
direct comparison of the WBL clusters with the Yerkes (AWM, MKW, \&
WP) poor clusters.  We list every Yerkes poor cluster along with any
corresponding WBL cluster.  If there is no WBL poor cluster associated
with a Yerkes cluster, we list a note detailing the reason it
was not included (see notes at the end of the table).

\begin{table}
\dummytable\label{tbl-4}
\end{table}

\section{Discussion}

Our catalog of poor clusters contains galaxy associations that span
several orders of magnitude in the cluster mass function---from very
poor systems of only three Zwicky galaxies, up to and including the
nearby Abell clusters.  Because the number of cluster galaxies listed
in Tables 2 and 3 is the number of galaxies brighter than
$m_{pg}=15.7$, it is not necessarily the total membership; fainter
galaxies may also be part of these clusters.  Further observations are
needed to characterize each WBL cluster fully in terms of membership,
galaxy morphology, velocity dispersion, etc.  From an analysis of a
number of the MKW and AWM clusters, Bahcall (1980) showed that these
poor clusters are just lower-richness extrapolations of Abell clusters
in terms of galaxy richness, galaxy density, and spiral fraction (see
also \cite{bha81}).  Other subsamples of the WBL catalog indicate
similar results in terms of velocity dispersion (\cite{led96};
\cite{bee95}), radio source population (\cite{bur87}; \cite{doe95}),
and x-ray properties (\cite{bur96}; \cite{doe95}; \cite{pri91}).

The WBL catalog covers the entire sky above $-3^{\circ}$ declination,
or approximately 52\% of the sky.  The locations of all clusters are
shown as dots in Figure 1.  Also plotted in Fig. 1 are circles
representing the Zwicky clusters (near and medium distant classes)
containing one or more WBL clusters.  Figure 1 illustrates the
interrelatedness of the Zwicky and WBL clusters; most (469 of 732) WBL
clusters are high density galaxy concentrations within the lower
density Zwicky cluster contours. Figure 2 shows the locations of WBL
poor clusters within the Zwicky clusters.  Although the number density
of WBL clusters peaks at small radii within the Zwicky contours, most
WBL clusters ($\approx$65\%) are located beyond $\frac{1}{2}$ of a
Zwicky radius.  This indicates subclustering of galaxies within many
Zwicky clusters.  Subclustering is also displayed in Figure 3, which
is a histogram of the number of WBL poor clusters contained within
each Zwicky contour.  Of the 245 Zwicky clusters that contain WBL poor
clusters, 97 ($\approx$40\%) contain multiple poor clusters.

Some (263) WBL clusters are found outside Zwicky clusters, as more
isolated galaxy associations.  In fact, there appear to be entire
regions of the sky where WBL poor clusters are found, yet are devoid
of Zwicky clusters (e.g. near $11^h$ RA, $+10^{\circ}$ Dec.---see
Fig. 1).  Additionally, many (274/504 $\approx$54\%) of the near
Zwicky clusters do not contain WBL clusters.  These are generally very
loose galaxy associations and are not detected at $\sigma_{21}$.
Using a lower $\sigma_g$ (such as $10^{1/3}$) one would detect many
more of these Zwicky clusters.  Also, the volume limit of the near
Zwicky clusters extends beyond the limiting redshift ($z$=0.03) to
which the WBL catalog is complete (see below).

Since the TG algorithm has no upper limit on its richness criterion,
the nearby Abell clusters with $\delta > -3^{\circ}$ are also found in
our catalog.  This was expected and desired, since we wished not only
to create a uniform sample of poor clusters but to explore their
relation to the rich clusters as well.  All distance class (DC) 0 and
1 clusters (a total of 19), and three of the five DC 2 clusters are
detected by the algorithm.  The two missing DC 2 Abell clusters have
galaxy densities just below the $\sigma_{21}$ detection threshold.  We
also detect nearly half of the DC 3 Abell clusters with $m_{10}<15.3$.
Of the 93 Abell clusters with DC$\leq$3 and $\delta > -3^{\circ}$, 45
($\approx$48\%) contain WBL clusters.  A distribution of the number
density of WBL clusters within Abell clusters is shown in Figure 4,
which clearly shows that, unlike the Zwicky clusters, the WBL catalog
finds the cores of Abell clusters.  This is expected, since Abell
clusters are generally more compact than most Zwicky clusters.
However, there is still evidence of possible subclustering, as a
significant fraction (35\%) of the Abell clusters coincident with our
catalog contain multiple WBL clusters (Figure 5).

As expected, the WBL catalog also has significant overlap with other
group catalogs, such as the HCG and CfA groups, but due to the
differing selection algorithms there is no direct correlation.
Although the cross-identifications are not listed in Table 2, the WBL
catalog also detects a number of TG groups, since the same algorithm
was used to find both sets of poor clusters.  TG searched to much
lower galaxy densities ($\rm 10^{2/3}$), and used a brighter magnitude
cutoff ($ m_{pg} < 14$), therefore focusing on very nearby, looser
galaxy associations in the north galactic cap ($b >40^{\circ}$,
$\delta > 0^{\circ}$).

The completeness of the entire WBL catalog is difficult to determine.
Certainly, the portions of the catalog at very low galactic latitudes
($|b|<30^{\circ}$) are incomplete due to obscuration by the galactic
plane.  Also, since there is no redshift information for a large
number of the galaxies in our catalog, a discussion of completeness,
and the frequency of projection effects, can only be approximate (see
\cite{bur96} for a discussion of projection effects in a subsample of
the catalog).  For poor clusters without well determined mean
redshifts, we estimated redshifts based on the magnitude of the
brightest cluster member ($m_1$).  We have used poor clusters from
Table 2, with no redshift notes (column 8), and thus reasonably well
determined redshifts, to calibrate the $m_1 - z$ relation for the
remaining poor clusters.  Included in this calibration are poor clusters
with greater than 7 reported redshifts, containing a $iii$ in column
8, which are certainly physical systems along with a few outlying
galaxies.  This relation, using 394 poor clusters, is shown in Figure
6.  From the redshifts in Table 2 combined with the estimated
redshifts for all other clusters, we have determined the volume
density of the WBL catalog for $|b|>30^{\circ}$.  In figure 7 we show
the volume density in redshift bins of $\Delta z=0.005$.  Based on
this analysis, this sample is nearly volume-limited over the redshift
range $0.01 < z < 0.03$.  A similar result was found for a smaller
sample of WBL clusters in Ledlow et al. (1996).  We are employing the
volume-limited portion of the WBL catalog ($|b|>30^{\circ},
0.01<z<0.03$) for an extensive x-ray, optical, and radio study of poor
clusters.

\section{Conclusions}

We present a catalog of 732 optically-selected poor clusters of
galaxies.  These WBL poor clusters were identified as galaxy surface
density enhancements in the CGCG, which cataloged galaxies brighter
than photographic magnitude 15.7, and declination greater than -3
degrees.  The WBL catalog covers a wide range of cluster richness,
from very poor systems containing only a few galaxies, through the
Yerkes poor clusters, and including many nearby Zwicky and Abell
clusters.  Previous analyses of the WBL clusters show a number of them
contain an x-ray bright intracluster medium (\cite{bur96};
\cite{pri91}) and interesting extragalactic radio morphologies
(\cite{bur87}; \cite{doe95}) similar to rich clusters. WBL clusters
with $|b|>30^{\circ}$ are nearly volume-limited in the redshift range
$0.01<z<0.03$ (Fig. 7), producing a subsample of $\approx$300 systems
which is ideal for studying the properties of poor clusters.  Taken in
its entirety, the WBL catalog covers over three orders of magnitude in
the cluster mass function ({\cite{bur96}) and provides an excellent
sample with which to study the formation and evolution of a wide range
of galaxy associations.

\acknowledgments

This work was completed with the assistance of NSF grant AST-986039
and NASA grant NAG5-6772 to J.O.B. \& M.B.  SPB wishes to acknowledge
the Kentucky Space Grant Consortium for its support, including
subgrant WKU 521781-97-02.  PLB's contribution was supported by the
NRAO Summer Student Program.

Shortly before this paper was submitted, the lead author, Dr. Richard
A. White, passed away following a long and valiant struggle with
leukemia.  We, his coauthors, friends, and colleagues, wish to
dedicate this paper to Richard.  His love for and devotion to
astronomy continue to inspire us all.

\newpage
\begin{deluxetable}{lrrrrrrrrrr}
\tablenum{1}
\tablecaption{Poor Cluster Statistics at Various Surface Densities \label{tbl-1}}
\tablewidth{0pt}
\tablehead{
\colhead{Galaxy Surface Density} & \colhead{$N_3$\tablenotemark{a}} & \colhead{$N_4$\tablenotemark{a}} & \colhead{$N_5$\tablenotemark{a}} & \colhead{$N_6$\tablenotemark{a}} & \colhead{$N_7$\tablenotemark{a}} & \colhead{$N_8$\tablenotemark{a}} & \colhead{$N_9$\tablenotemark{a}} & \colhead{$N_{>9}$\tablenotemark{a}} & \colhead{$N_{tot}$\tablenotemark{b}} & \colhead{$G_{tot}$\tablenotemark{c}}}
\startdata
$\sigma_{21}$  & 446 & 127 & 54 & 31 & 25 & 10 & 9 & 30 & 732 & 3324 \nl
$\sigma_{46}$  & 262 & 61  & 23 & 15 & 6  & 6  & 6 & 12 & 391 & 1691 \nl

\enddata
\tablenotetext{a}{$N_i$ = number of poor clusters containing $i$ Zwicky galaxies}
\tablenotetext{b}{$N_{tot}$ = number of poor clusters with $\geq$3 Zwicky galaxies}
\tablenotetext{c}{$G_{tot}$ = number of Zwicky galaxies in poor clusters}
\end{deluxetable}

\newpage
\begin{deluxetable}{ccccrrccl}
\tablenum{3}
\tablewidth{0pc}
\tablecaption{Poor Cluster Galaxies (sample)}
\tablehead{
\colhead{Name} & \multicolumn{1}{c}{RA} &
\multicolumn{1}{c}{Dec} &
\colhead{$m_{pg}$} & \colhead{$N_{21}$} &
\colhead{$N_{46}$} & \colhead{$\sigma_{46}$ Fate} &
\colhead{$z$} & \multicolumn{1}{l}{Other names} \\
\colhead{} & \multicolumn{1}{c}{B1950} &
\multicolumn{1}{c}{B1950} &
\colhead{} & \colhead{} &
\colhead{} & \colhead{} &
\colhead{} & \colhead{} \\
\colhead{(1)} & \colhead{(2)} & \colhead{(3)} &
\colhead{(4)} & \colhead{(5)} & \colhead{(6)} &
\colhead{(7)} & \colhead{(8)} & \multicolumn{1}{l}{(9)}
}
\startdata
WBL214$-$001 & 09 15 06 & $\rm +20 \ 41$ & 15.6 &   0 &   0 & \nodata & 0.03031 & \nodata \nl
WBL214$-$002 & 09 15 24 & $\rm +20 \ 35$ & 15.7 &   3 &   1 & A & 0.03185 & \nodata \nl
WBL214$-$003 & 09 15 30 & $\rm +20 \ 28$ & 15.5 &   0 &   0 & \nodata & 0.02788 & \nodata \nl
WBL214$-$004 & 09 15 36 & $\rm +20 \ 37$ & 15.7 &   1 &   1 & A & 0.03193 & \nodata \nl
       &          & $\rm     \   $ &      &     &     &   &         & \nl
WBL215$-$001 & 09 16 18 & $\rm +06 \ 06$ & 14.9 &   1 &   1 & A & \nodata & \nodata \nl
WBL215$-$002 & 09 16 30 & $\rm +06 \ 06$ & 15.5 &   2 &   1 & A & \nodata & \nodata \nl
WBL215$-$003 & 09 16 54 & $\rm +06 \ 06$ & 15.7 &   1 &   0 & \nodata & \nodata & \nodata \nl
       &          & $\rm     \   $ &      &     &     &   &         & \nl
WBL216$-$001 & 09 15 18 & $\rm +33 \ 46$ & 15.6 &   0 &   0 & \nodata & 0.02401 & \nodata \nl
WBL216$-$002 & 09 15 18 & $\rm +34 \ 04$ & 15.6 &   1 &   1 & A & 0.02220 & \nodata \nl
WBL216$-$003 & 09 15 30 & $\rm +34 \ 04$ & 15.6 &   1 &   1 & A & 0.02414 & \nodata \nl
WBL216$-$004 & 09 16 12 & $\rm +34 \ 04$ & 15.6 &  10 &   1 & B & 0.02341 & NGC2827  IC2460 \nl
WBL216$-$005 & 09 16 12 & $\rm +34 \ 13$ & 15.7 &   0 &   0 & \nodata & 0.02366 & NGC2823 UGC04935 \nl
WBL216$-$006 & 09 16 18 & $\rm +33 \ 57$ & 15.3 &  12 &   0 & C & 0.02659 & NGC2825 \nl
WBL216$-$007 & 09 16 24 & $\rm +33 \ 50$ & 14.6 &   0 &   0 & \nodata & 0.02473 & NGC2826 UGC04939 \nl
WBL216$-$008 & 09 16 30 & $\rm +34 \ 05$ & 15.7 &  10 &   1 & B & 0.02166 & NGC2828 \nl
WBL216$-$009 & 09 16 42 & $\rm +33 \ 57$ & 15.4 &   8 &   3 & C & 0.02037 & NGC2830 UGC04941 \nl
WBL216$-$010 & 09 16 48 & $\rm +33 \ 58$ & 13.3 &   8 &   2 & C & 0.02096 & NGC2832 UGC04942 \nl
WBL216$-$011 & 09 16 54 & $\rm +34 \ 08$ & 15.6 &   1 &   0 & \nodata & 0.02475 & NGC2833 \nl
WBL216$-$012 & 09 17 00 & $\rm +33 \ 55$ & 15.6 &   8 &   2 & C & 0.02351 & NGC2834 \nl
WBL216$-$013 & 09 17 36 & $\rm +33 \ 52$ & 15.3 &   0 &   0 & \nodata & 0.02743 & NGC2839 \nl
       &          & $\rm     \   $ &      &     &     &   &         & \nl
WBL217$-$001 & 09 17 18 & $\rm +01 \ 09$ & 15.4 &   2 &   1 & A & 0.01778 & \nodata \nl
WBL217$-$002 & 09 17 24 & $\rm +01 \ 08$ & 15.5 &   2 &   1 & A & 0.01724 & \nodata \nl
WBL217$-$003 & 09 17 30 & $\rm +01 \ 15$ & 14.1 &   0 &   0 & \nodata & 0.01718 & UGC04956 \nl
       &          & $\rm     \   $ &      &     &     &   &         & \nl
WBL218$-$001 & 09 17 06 & $\rm +64 \ 28$ & 14.0 &   2 &   2 & A & 0.00544 & NGC2814 \nl
WBL218$-$002 & 09 17 30 & $\rm +64 \ 27$ & 15.1 &   2 &   2 & A & 0.00512 & NGC2820  IC2458 \nl
WBL218$-$003 & 09 17 48 & $\rm +64 \ 29$ & 13.1 &   2 &   2 & A & 0.00525 & NGC2820 UGC04961 \nl
\enddata
\end{deluxetable}

\newpage
\begin{deluxetable}{ll}
\tablecaption{Old Poor Cluster Designations Used in the Literature \label{tbl-4}
}
\tablewidth{0pt}
\tablenum{4}
\tablehead{
\colhead{Old Name} & \colhead{New Name}
}
\startdata
AWM1 & WBL213 \nl
AWM2 & WBL389 \nl
AWM3 & WBL509 \nl
AWM4 & a,b \nl
AWM5 & WBL625 \nl
AWM6 & b \nl
AWM7 & WBL088 \nl
 MKW01 & WBL248 \nl
 MKW02 & WBL276 \nl
 MKW03 & WBL360 \nl
 MKW04 & WBL372 \nl
 MKW05 & WBL483 \nl
 MKW06 & WBL498 \nl
 MKW07 & WBL514 \nl
 MKW08 & WBL518 \nl
 MKW09 & WBL574 \nl
 MKW10 & WBL350 \nl
 MKW11 & WBL447 \nl
 MKW12 & WBL486 \nl
MKW01s & WBL217 \nl
MKW02s & WBL272 \nl
MKW03s & WBL564 \nl
MKW04s & WBL378 \nl
WP01 & WBL007 \nl
WP02 & c \nl
WP03 & c \nl 
WP04 & WBL042 \nl
WP05 & c \nl
WP06 & WBL061 \nl
WP07 & c \nl
WP08 & c \nl
WP09 & c \nl
WP10 & WBL133 \nl
WP11 & c \nl
WP12 & WBL248 \nl
WP13 & WBL280 \nl
WP14 & WBL288 \nl
WP15 & c \nl
WP16 & WBL350 \nl
WP17 & WBL360 \nl
WP18 & b \nl
WP19 & WBL368 \nl
WP20 & WBL374 \nl
WP21 & WBL374 \nl
WP22 & WBL430 \nl
WP23 & c \nl
WP24 & c \nl
WP25 & b \nl
WP26 & WBL457 \nl
WP27 & b \nl
WP28 & WBL477 \nl
WP29 & WBL478 \nl
WP30 & WBL486 \nl
WP31 & b \nl
WP32 & WBL548 \nl
WP33 & WBL593 \nl
WP34 & WBL611 \nl
WP35 & WBL625 \nl
WP36 & c \nl
WP37 & c \nl
WP38 & WBL690 \nl
WP39 & WBL700 \nl
WP40 & WBL706 \nl
WP41 & b \nl
WP42 & c \nl
   K103 & WBL312 \nl
   K131 & WBL472 \nl
   K140 & WBL499 \nl
   K179 & WBL226 \nl
    K30 & WBL288 \nl
   K315 & WBL438 \nl
    K36 & WBL323 \nl
N34-169 & WBL210 \nl
N34-170 & WBL234 \nl
N34-171 & WBL622 \nl
N34-172 & WBL248 \nl
N34-173 & WBL650 \nl
N34-175 & WBL636 \nl
N45-273 & WBL205 \nl
N45-280 & WBL222 \nl
N45-281 & WBL223 \nl
N45-285 & WBL230 \nl
N45-300 & WBL257 \nl
N45-302 & WBL269 \nl
N45-309 & WBL550 \nl
N45-342 & WBL206 \nl
N45-346 & WBL231 \nl
N45-348 & WBL234 \nl
N45-351 & WBL245 \nl
N45-353 & WBL256 \nl
N45-356 & WBL551 \nl
N45-358 & WBL565 \nl
N45-360 & WBL574 \nl
N45-361 & WBL588 \nl
N45-363 & WBL597 \nl
N45-365 & WBL622 \nl
N45-366 & WBL219 \nl
N45-367 & WBL243 \nl
N45-368 & WBL248 \nl
N45-371 & WBL322 \nl
N45-376 & WBL210 \nl
N45-377 & WBL232 \nl
N45-379 & WBL579 \nl
N45-381 & WBL554 \nl
N45-382 & WBL573 \nl
N45-383 & WBL619 \nl
N45-384 & WBL224 \nl
N45-388 & WBL213 \nl
N45-389 & WBL612 \nl
N45-391 & WBL566 \nl
N56-296 & WBL298 \nl
N56-297 & WBL299 \nl
N56-306 & WBL336 \nl
N56-308 & WBL346 \nl
N56-329 & WBL522 \nl
N56-332 & WBL533 \nl
N56-336 & WBL539 \nl
N56-338 & WBL542 \nl
N56-339 & WBL543 \nl
N56-343 & WBL548 \nl
N56-355 & WBL287 \nl
N56-359 & WBL341 \nl
N56-361 & WBL497 \nl
N56-365 & WBL537 \nl
N56-366 & WBL588 \nl
N56-367 & WBL271 \nl
N56-368 & WBL279 \nl
N56-369 & WBL288 \nl
N56-371 & WBL355 \nl
N56-373 & WBL576 \nl
N56-374 & WBL577 \nl
N56-381 & WBL524 \nl
N56-385 & WBL540 \nl
N56-387 & WBL270 \nl
N56-388 & WBL280 \nl
N56-391 & WBL312 \nl
N56-392 & WBL360 \nl
N56-393 & WBL258 \nl
N56-394 & WBL514 \nl
N56-395 & WBL518 \nl
N56-396 & WBL324 \nl
N67-238 & WBL419 \nl
N67-244 & WBL303 \nl
N67-245 & WBL304 \nl
N67-300 & WBL358 \nl
N67-309 & WBL508 \nl
N67-310 & WBL527 \nl
N67-311 & WBL326 \nl
N67-312 & WBL350 \nl
N67-317 & WBL435 \nl
N67-318 & WBL472 \nl
N67-320 & WBL318 \nl
N67-321 & WBL325 \nl
N67-322 & WBL334 \nl
N67-323 & WBL430 \nl
N67-325 & WBL493 \nl
N67-326 & WBL509 \nl
N67-328 & WBL311 \nl
N67-329 & WBL450 \nl
N67-330 & WBL397 \nl
N67-333 & WBL428 \nl
N67-335 & WBL372 \nl
N67-336 & WBL486 \nl
N79-207 & WBL335 \nl
N79-220 & WBL373 \nl
N79-238 & WBL419 \nl
N79-266 & WBL375 \nl
N79-268 & WBL395 \nl
N79-270 & WBL438 \nl
N79-276 & WBL478 \nl
N79-278 & WBL343 \nl
N79-280 & WBL356 \nl
N79-281 & WBL381 \nl
N79-282 & WBL385 \nl
N79-283 & WBL394 \nl
N79-284 & WBL415 \nl
N79-286 & WBL475 \nl
N79-290 & WBL354 \nl
N79-292 & WBL404 \nl
N79-293 & WBL408 \nl
N79-295 & WBL428 \nl
N79-296 & WBL447 \nl
N79-297 & WBL477 \nl
N79-298 & WBL368 \nl
N79-298 & WBL368 \nl
N79-299 & WBL374 \nl
N79-299 & WBL374 \nl
S34-087 & WBL008 \nl
S34-088 & WBL017 \nl
S34-090 & WBL020 \nl
S34-091 & WBL026 \nl
S34-093 & WBL062 \nl
S34-103 & WBL713 \nl
S34-106 & WBL731 \nl
S34-110 & WBL009 \nl
S34-111 & WBL025 \nl
S34-112 & WBL724 \nl
S34-113 & WBL038 \nl
S34-115 & WBL007 \nl
S49-122 & WBL015 \nl
S49-123 & WBL017 \nl
S49-125 & WBL044 \nl
S49-126 & WBL050 \nl
S49-127 & WBL075 \nl
S49-128 & WBL083 \nl
S49-130 & WBL686 \nl
S49-132 & WBL698 \nl
S49-134 & WBL706 \nl
S49-137 & WBL717 \nl
S49-138 & WBL726 \nl
S49-139 & WBL727 \nl
S49-140 & WBL061 \nl
S49-141 & WBL063 \nl
S49-142 & WBL099 \nl
S49-143 & WBL695 \nl
S49-144 & WBL702 \nl
S49-145 & WBL066 \nl
S49-146 & WBL688 \nl
S49-147 & WBL009 \nl
\enddata
\tablenotetext{a}{Cluster galaxies were too faint to be included in the WBL catalog.}
\tablenotetext{b}{Cluster is too diffuse to meet the galaxy density criterion ofthe WBL catalog.}
\tablenotetext{c}{Cluster is south of $\rm -3^{\circ}$ declination, and therefore not included in the WBL catalog.}
\end{deluxetable}

\newpage

\begin{figure}
\plotone{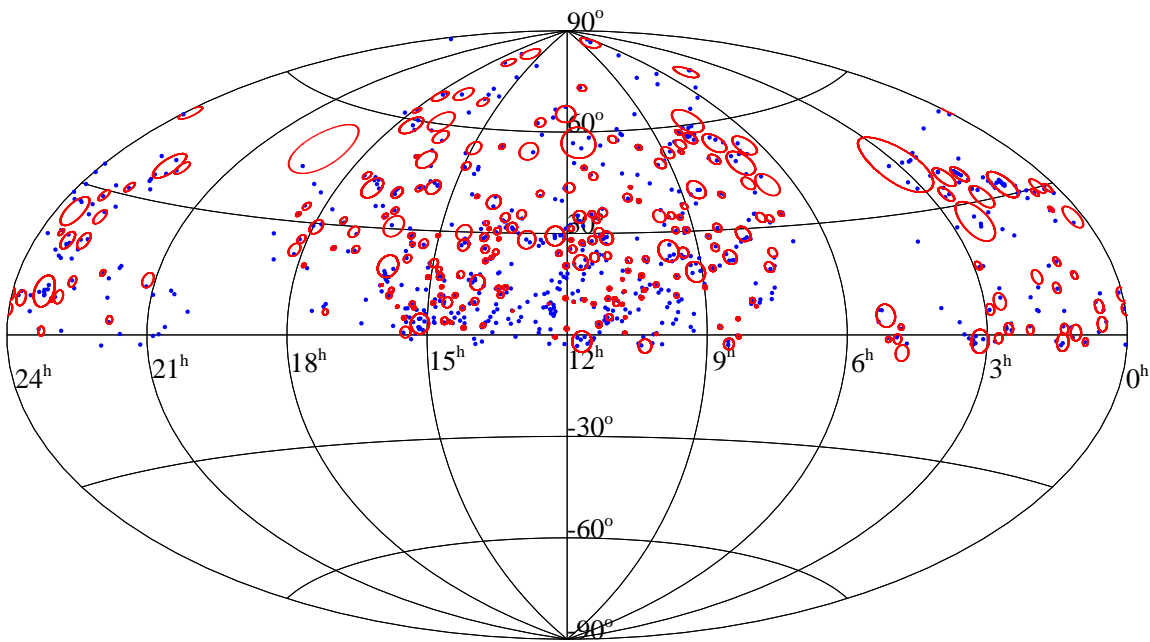}
\caption{The location on the sky of all 732 WBL clusters (blue dots), and all Zwicky clusters (red circles) that contain a poor cluster.  The size of the circle corresponds to the Zwicky radius from the CGCG (Table 2, column 9).  The empty regions at positive declination are due to obscuration by the galactic plane. \label{fig1}}
\end{figure}

\begin{figure}
\plotone{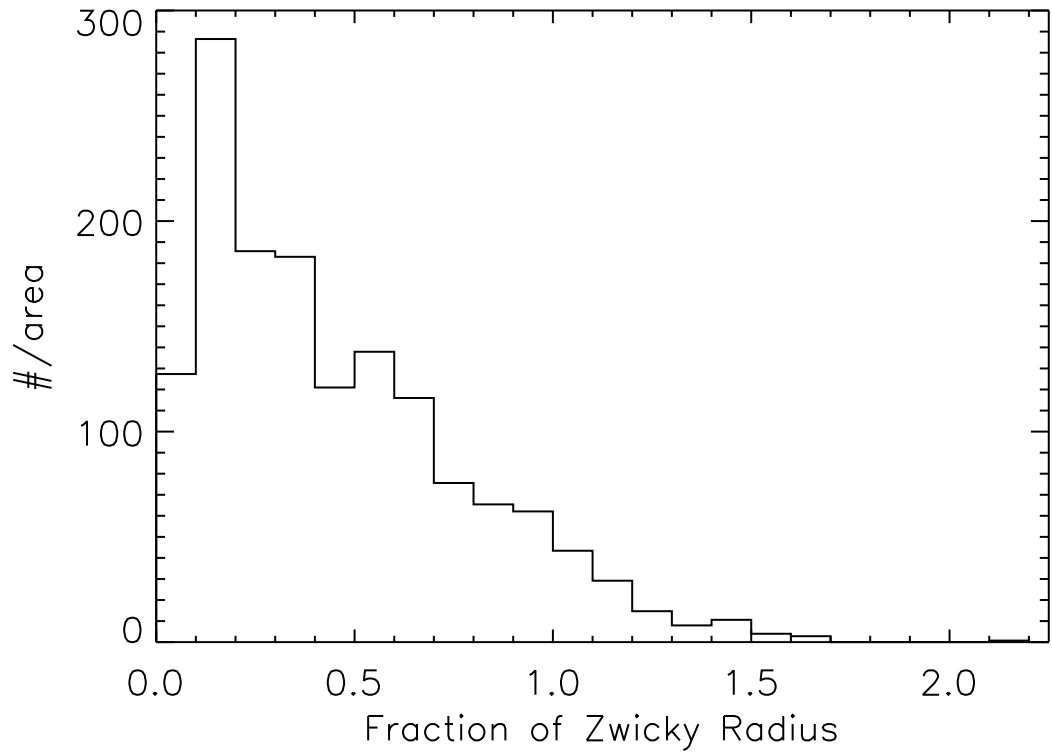}
\caption{The surface number density of all WBL clusters within Zwicky clusters, as a function of the Zwicky radius.  WBL poor clusters tend to avoid the centers of Zwicky clusters. \label{fig2}}
\end{figure}

\begin{figure}
\plotone{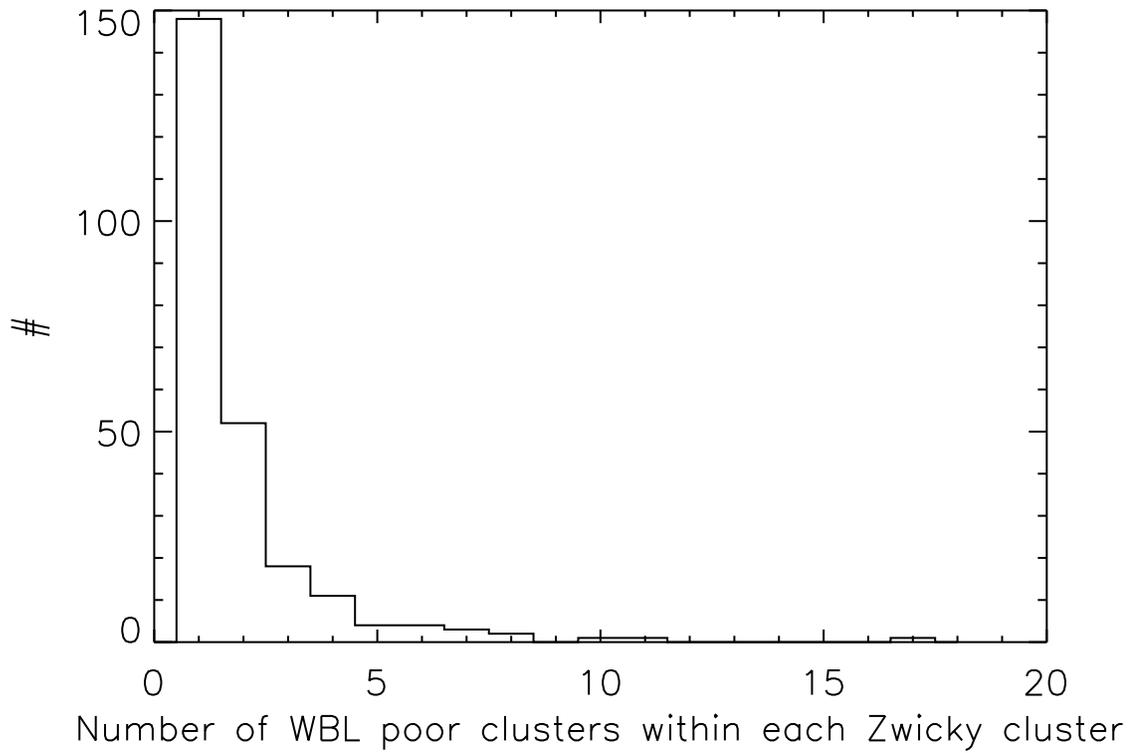}
\caption{The number of WBL clusters within Zwicky clusters.  Approximately 40\% of Zwicky clusters which possess WBL clusters contain multiple poor clusters. \label{fig3}}
\end{figure}

\begin{figure}
\plotone{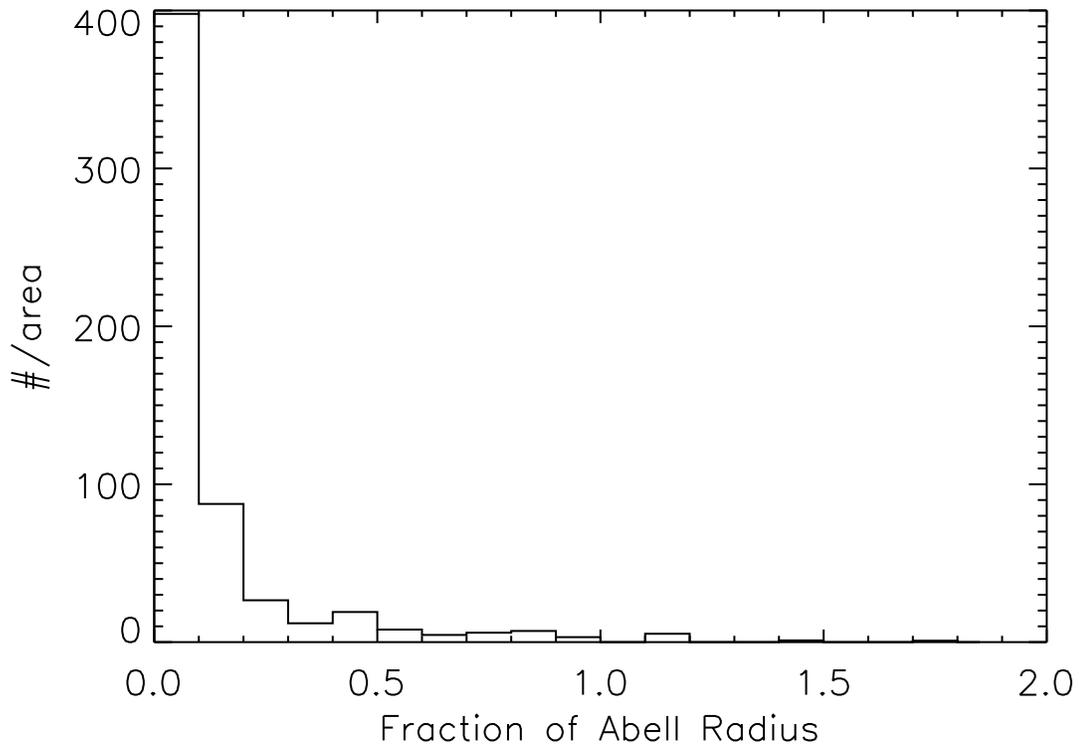}
\caption{The surface number density of all WBL clusters within Abell clusters, as a function of the Abell radius.  Most WBL poor clusters are located near the centers of Abell clusters. \label{fig4}}
\end{figure}

\begin{figure}
\plotone{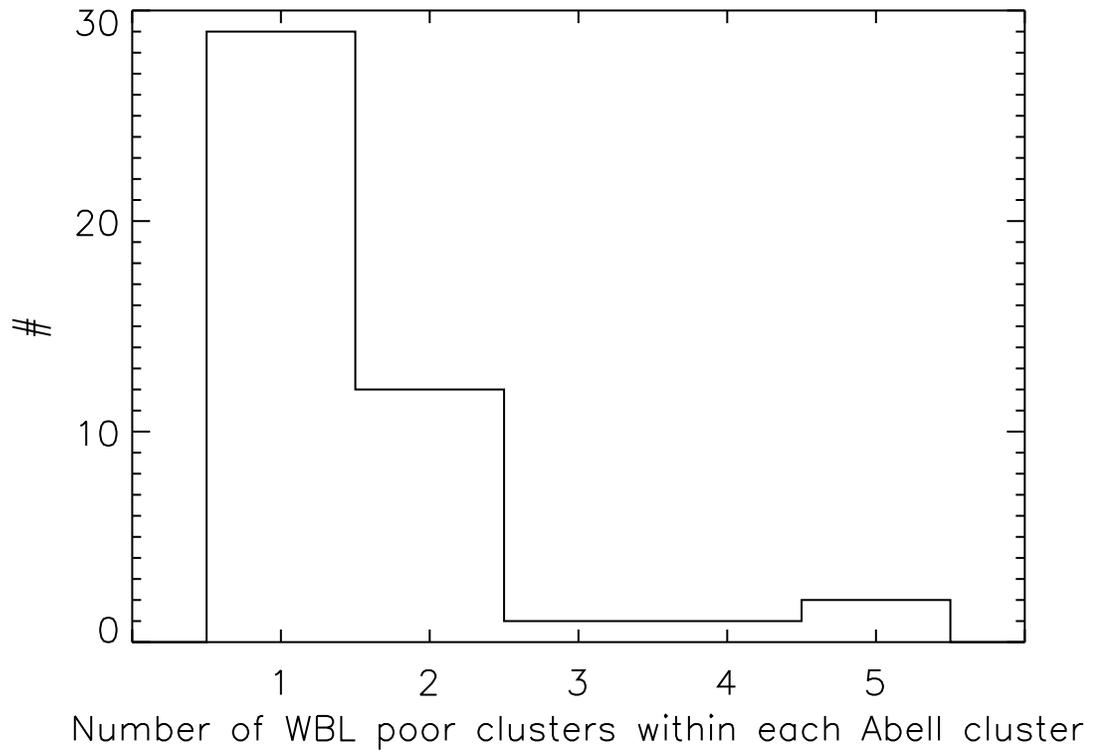}
\caption{The number of WBL clusters within Abell clusters.  Approximately 35\% of Abell clusters which possess WBL clusters contain multiple poor clusters. \label{fig5}}
\end{figure}

\begin{figure}
\plotone{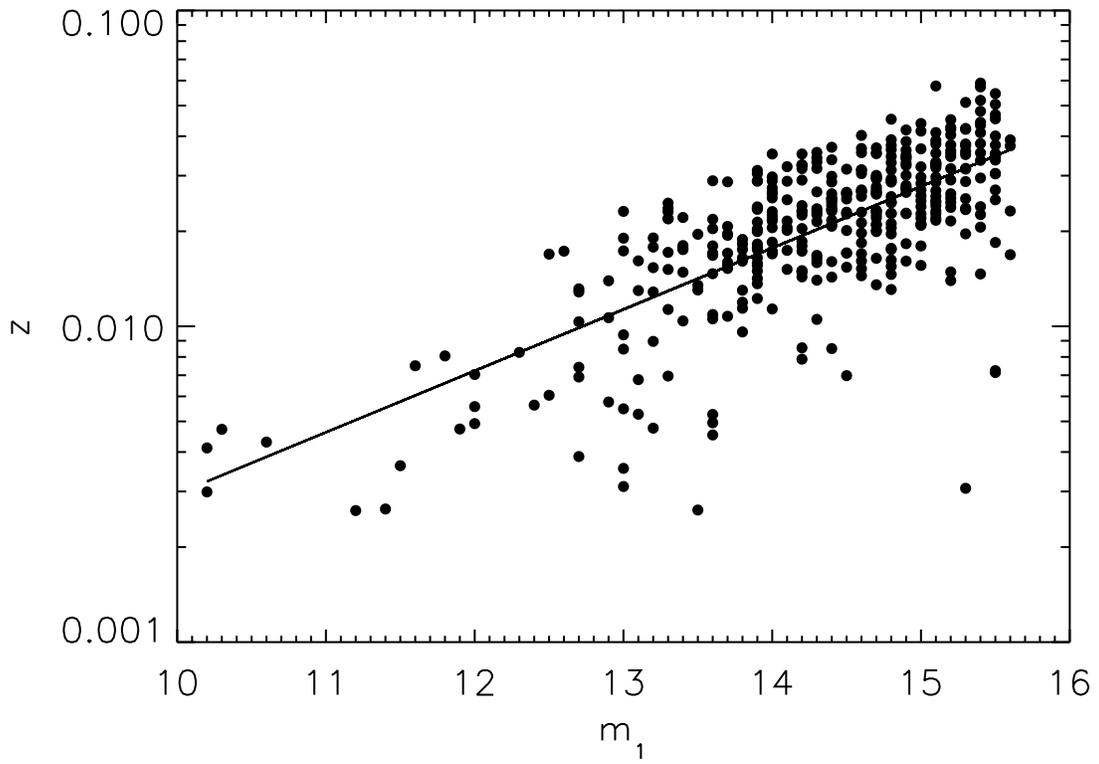}
\caption{The relation used to estimate redshifts, $z$, from the apparent magnitude of the brightest cluster member, $m_1$. \label{fig6}}
\end{figure}

\begin{figure}
\plotone{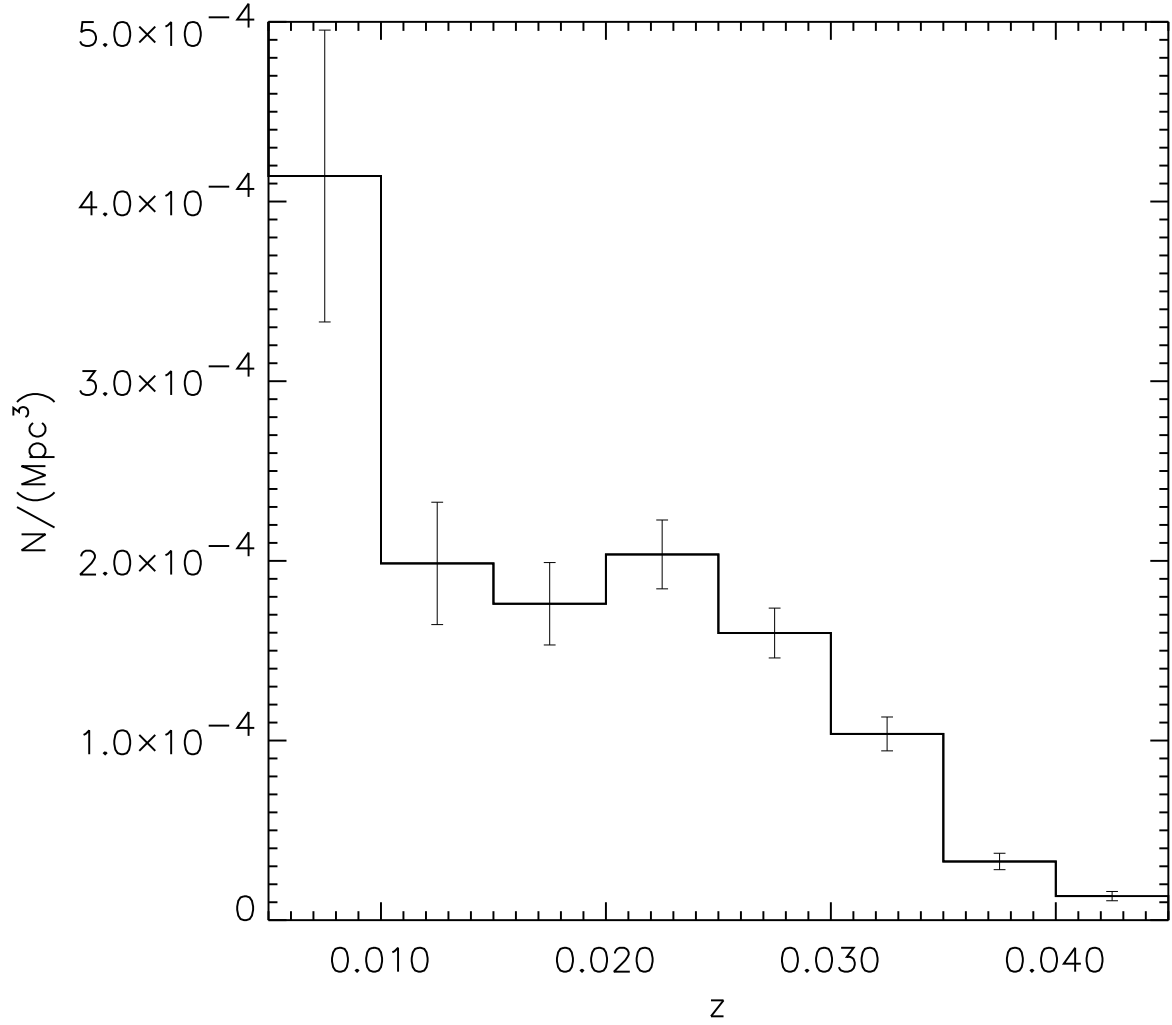}
\caption{The volume density of WBL poor clusters in the galactic latitude range $|b| > 30^{\circ}$.  We judge this poor cluster sample to be volume-limited in the redshift range $0.01 < z < 0.03$. Error bars represent Poisson errors in each bin. \label{fig7}}
\end{figure}

\end{document}


